\documentclass[aps,pre,twocolumn,floatfix,superscriptaddress,showpacs,showkeys]{revtex4-1}
\usepackage{epsf,amsmath,amssymb,verbatim,color,multirow,pifont,mathrsfs,soul,amsthm, wasysym}
\usepackage{graphicx}
\usepackage{upgreek}
\usepackage[us,12hr]{datetime}

\begin{document}
\today

\title{Concurrent formation of nearly synchronous clusters in each intertwined cluster set with parameter mismatches}
\author{Young Sul Cho}\email{yscho@jbnu.ac.kr}
\affiliation{Department of Physics, Chonbuk National University, Jeonju 54896, Korea}
\affiliation{Department of Physics, Research Institute of Physics and Chemistry, Chonbuk National University, Jeonju 54896, Korea}

\begin{abstract}
Cluster synchronization is a phenomenon in which oscillators in a given network are partitioned into synchronous clusters. 
As recently shown, diverse cluster synchronization patterns can be found using network symmetry when the oscillators are identical.
For such symmetry-induced cluster synchronization patterns, subsets called intertwined clusters can exist, in which every cluster in the same subset should synchronize or desynchronize concurrently.
In this work, to reflect the existence of noise in real systems, we consider networks composed of nearly identical oscillators. 
We show that every cluster in the same intertwined cluster set is nearly synchronized concurrently when the nearly synchronous 
state of the set is stable. 
We also consider an extreme case where only one cluster of an intertwined cluster set 
is composed of nearly identical oscillators while every other cluster in the set is composed of identical oscillators. 
In this case, deviation from the synchronous state of every cluster in the same set increases linearly 
with the magnitude of parameter mismatch within the cluster
of nearly identical oscillators. We confirm these results by numerical simulation.
\end{abstract}

\maketitle

\section{introduction}

Synchronization is a phenomenon in which the states of interacting oscillators evolve with
the same rate of change~\cite{strogatz_sync, kurts_sync}.
These collective behaviors can be observed in a variety of real systems, such as flashing fireflies~\cite{walker_firefly}, firing neurons in the brain~\cite{sync_neural, sync_neural2}, electric power grids~\cite{motter_powergrid, kuramoto_powergrid},
and others~\cite{arenas_review}. 
Within this phenomenon, cluster synchronization (CS), which is a partition of oscillators in a network into synchronized subsets (clusters), 
has been widely studied~\cite{cluster_sync1, cluster_sync2, cluster_sync3, equitable_partition}.

Network symmetry is a permutation of oscillators that conserves
the dynamical system~\cite{symmetry}. Diverse relations between network symmetry and synchronization
have been discovered, including remote synchronization~\cite{remote_prl_2013}, 
isolated desynchronization~\cite{pecora_ncomm_2014}, and asymmetry-induced synchronization~\cite{aisync_takashi,aisync_yuanzhao}.
Recently, it has been reported that diverse CS patterns can be found using the symmetry
of a network composed of {\it identical} oscillators~\cite{pecora_ncomm_2014, pecora_sciadv_2016}. 
For an arbitrary symmetry-induced CS pattern, clusters can be divided into subsets where stability is coupled between all clusters within the subset, but decoupled from the clusters outside of the subset~\cite{pecora_ncomm_2014, yscho_prl_2017}.
Each subset is called a set of {\it intertwined clusters} if the number of clusters in the subset is larger than one;
otherwise, it is called a {\it non-intertwined cluster}~\cite{pecora_ncomm_2014}. 
Therefore, every cluster in the same set of intertwined clusters is either stable or unstable at the same time. 
If one or more clusters in each subset are stable,
then cluster synchronization of the subset is observable.

However, oscillators in real systems cannot be exactly identical due to noise. 
Using {\it nearly identical} oscillators with small parameter mismatches
\cite{nearly_identical_origin, takashi_nearly_identical, sorrentino_nearly_identical, nearly_identical_2012},
it has been shown that a non-intertwined cluster can be nearly synchronous if it is stable~\cite{sorrentino2016}.
In the current paper, we extend this result to the case of intertwined clusters. 
We first establish the condition for stable, nearly synchronous CS of each intertwined cluster set, 
and then demonstrate that this phenomenon can be
observed if the set is stable. We believe that this result can explain diverse nearly synchronous
CS patterns including twisted states in square and cubic lattices~\cite{twisted_2018}, as discussed in Sec.~\ref{sec:discussion}.

The rest of this paper is organized as follows. In Sec.~\ref{sec:model}, we describe the model
used in this study, and in Sec.~\ref{sec:background} we review previous studies
that are useful to understand the present work. In Sec.~\ref{sec:cs_intertwined},
we demonstrate a theoretical framework for the stable, nearly synchronous CS of each intertwined cluster set, and
in Sec.~\ref{sec:example} we test the framework with an example and confirm its validity.
We discuss the results in Sec.~\ref{sec:discussion} 
and provide details supporting our analysis in the Appendix.

\section{Model}
\label{sec:model}
In this section, we describe the dynamical system considered in this paper.
The model consists of $N$ number of oscillators that are connected with each other in a given network, with
the given network structure described by $N \times N$ adjacency matrix $\bold{A}$ 
whose element $A_{ij}=1$ if oscillators $i$ and $j$ are connected or $A_{ij}=0$ otherwise.
For simplicity, we only consider a bidirectional network (i.e. $\bold{A}$ is symmetric). 

The state of each oscillator $i$ at time $t$ is described by $n$-dimensional vector $\bold{x}_i(t) \in \mathbb{R}^n$.
The governing equation for $\bold{x}_i(t)$ is given by
\begin{equation}
\dot{\bold{x}}_i(t)=\bold{F}({\bold x}_i(t), {\boldsymbol \upmu}_i) + \sigma \sum_{j=1}^N A_{ij}{\bold H}(\bold{x}_j(t))
\label{eq:govern}
\end{equation}
for $1 \leq i, j \leq N$, where $q$-dimensional {\it time-independent} vector $\boldsymbol{\upmu}_i \in \mathbb{R}^q$ 
is the internal parameter of each oscillator $i$.
Here, ${\bold F}({\bold x}, \boldsymbol{\upmu})$ is a function for the dynamics of each oscillator 
when one is disconnected from all the others, while ${\bold H}({\bold x})$ is a
function for the interaction between connected oscillators. $\sigma \in \mathbb{R}$ is the global coupling strength.
We note that all the oscillators are identical if $\boldsymbol{\upmu}_i=\overline{\boldsymbol{\upmu}}$ for $\forall i$.

\section{Background}
\label{sec:background}
\subsection{Symmetry-induced CS patterns}
\label{sec:sics}

It has been shown that diverse CS patterns can be captured using the symmetry 
of a given network structure when the network is composed of identical oscillators
(i.e. $\boldsymbol{\upmu}_i=\overline{\boldsymbol{\upmu}}$ for $\forall i$)~\cite{pecora_ncomm_2014, pecora_sciadv_2016}. 
To describe network symmetry, automorphisms of the network have been used.  
An automorphism is a permutation $\pi$ of the oscillator set $\{i\}_{1 \leq i \leq N}$ 
that preserves the adjacency matrix such that $A_{ij}=A_{\pi(i)\pi(j)}$.
Then, the automorphism group of ${\bold A}$ denoted by $\textrm{Aut}(\bold{A})$ 
is the (mathematical) group consisting of all automorphisms of ${\bold A}$.

For each subgroup $G \leq \textrm{Aut}(\bold{A})$, the orbit of oscillator $i$ acted upon by $G$ is defined by
$\varphi(G, i)=\{\pi(i)|\pi \in G\}$. By the properties of a group, 
$\varphi(G, i) = \varphi(G, j)$ for $\forall{j \in \varphi(G, i)}$, which means that each oscillator 
belongs to a unique orbit of $G$. 
Therefore, each $G$ partitions the oscillators into associated orbits.
This partition can be a CS pattern of identical oscillators, as discussed below.

We consider the set of orbits $\{C_m\}_{1\leq m\leq M}$ given by subgroup $G$.
For an associated CS trajectory, $\{{\bold x}_i(t) = {\bold s}_m(t)~|~i \in C_m,~1 \leq m \leq M\}$, 
Eq.~(\ref{eq:govern}) for $\boldsymbol{\upmu}_i=\overline{\boldsymbol{\upmu}}$ can be reduced to 
{\it quotient network dynamics} such as
\begin{equation}
\dot{\bold s}_m(t) = {\bold F}({\bold s}_m(t), \overline{\boldsymbol{\upmu}}) + \sigma \sum_{m'=1}^M 
\widetilde{A}_{mm'} {\bold H}({\bold s}_{m'}(t))
\label{eq:govern_CS}
\end{equation}
for $1 \leq m,m' \leq M$, where quotient network adjacency matrix 
$\widetilde{A}_{mm'} = \sum_{j \in {C_{m'}}}A_{ij}$ for an arbitrary $i \in C_m$.
Here, it is guaranteed that $\widetilde{A}_{mm'}$ is the same regardless of $i \in C_m$ because 
all $i \in C_m$ receive the same input from every other cluster by symmetry~\cite{pecora_ncomm_2014}.
This means that CS trajectory evolves following Eq.~(\ref{eq:govern_CS}).
In principle, all symmetry-induced CS patterns can be found by investigating $\forall G \leq \textrm{Aut}(\bold{A})$~\cite{pecora_sciadv_2016}.
Moreover, we remark that multiple subgroups of $\textrm{Aut}(\bold{A})$ can be associated with the same CS pattern in general.

\subsection{Capturing intertwined cluster sets in a symmetry-induced CS pattern}
\label{sec:intertwined}

In this section, we review the method to capture non-intertwined clusters and intertwined cluster sets of an arbitrary 
symmetry-induced CS pattern, as reported in~\cite{yscho_prl_2017}.
Specifically, we consider network structure $\bold{A}$ and use ${\bold C}^{(G)}$ to denote the set of all nontrivial clusters (containing more than one oscillator) 
belonging to the CS pattern given by $G \leq {\textrm{Aut}}(\bold {A})$.

For a CS pattern given by $G \leq {\textrm{Aut}}(\bold {A})$, we first identify the non-intertwined clusters of ${\bold C}^{(G)}$. 
Each $C_m \in {\bold C}^{(G)}$ is a non-intertwined cluster if there exists at least one $G_1 \leq {\textrm {Aut}}({\bold A})$
that satisfies ${\bold C}^{(G_1)}=\{C_m\}$. In this manner, we can uniquely identify the set of all non-intertwined clusters of ${\bold C}^{(G)}$
which is denoted by ${\bold C}$.

For the other nontrivial clusters ${\bold C}^{(G)}-{\bold C}$, we then identify the intertwined cluster sets.
A subset ${\bold C}_1 \subseteq {\bold C}^{(G)}-{\bold C}$ is a set of intertwined clusters
if there exists at least one $G_2 \leq {\textrm {Aut}}({\bold A})$ that satisfies ${\bold C}_1 = {\bold C}^{(G_2)}$ 
and there is no $G_3 \leq {\textrm{Aut}}({\bold A})$ for which 
${\bold C}^{(G_3)}$ is a proper subset of ${\bold C}_1$.

\begin{figure}[t!]
\includegraphics[width=0.8\linewidth]{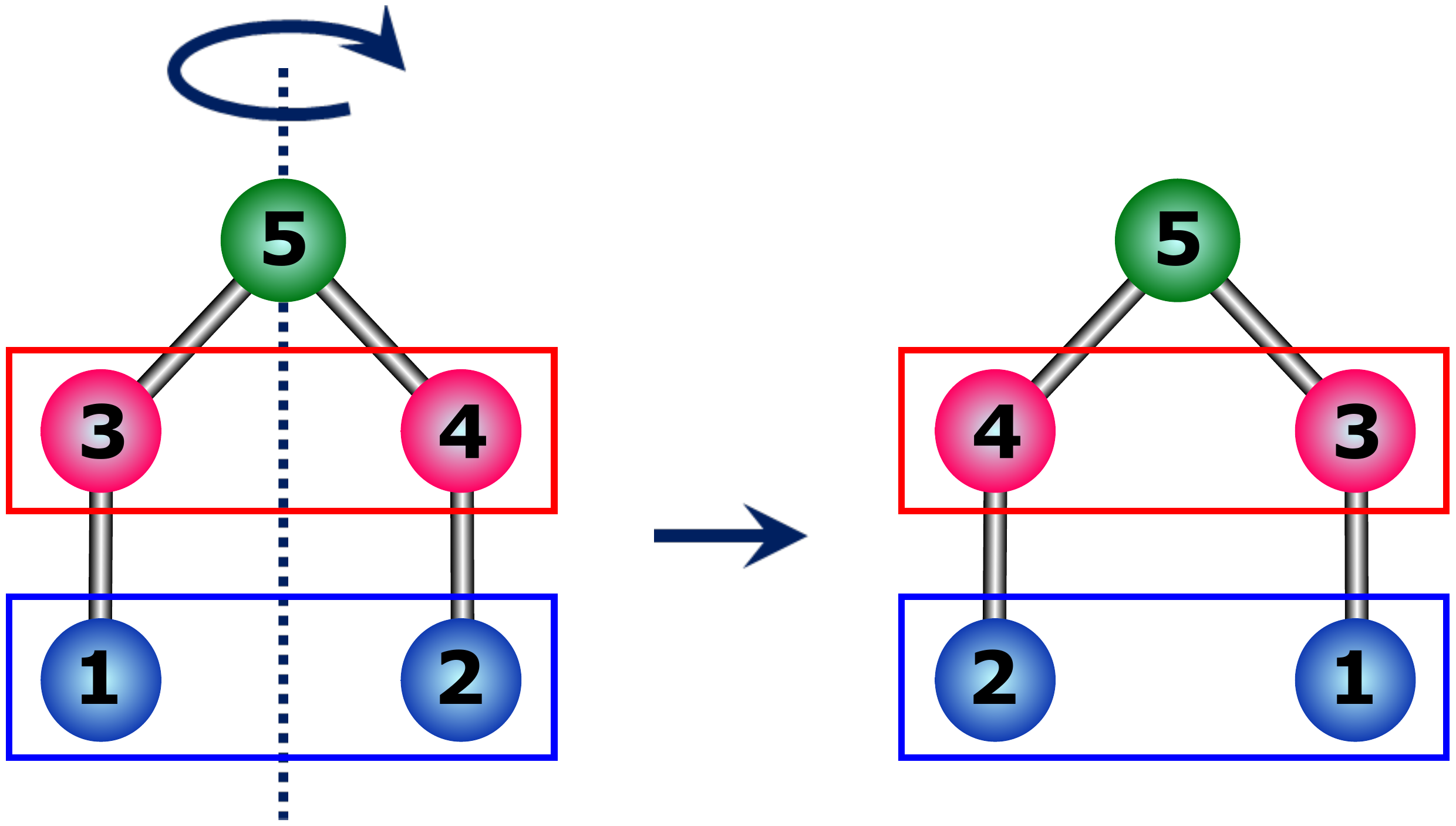}
\caption{Schematic diagram of a set of intertwined clusters 
$\{C_1,C_2\}$ within a CS pattern $\{C_1,C_2,C_3\}$, where
$C_1=\{1,2\}$, $C_2=\{3,4\}$, and $C_3=\{5\}$.
The two pairs of oscillators $C_1$ and $C_2$ should be permuted at the same time to conserve the adjacency matrix,
such that both clusters synchronize or desynchronize at the same time.
}   
\label{Fig:Intertwined_clusters_used}
\end{figure}

It has been shown that an arbitrary symmetry-induced CS pattern can be uniquely grouped into 
non-intertwined clusters and intertwined cluster sets~\cite{yscho_prl_2017};
computational codes for such grouping are presented in~\cite{github_2017}.
An example of a set of intertwined clusters is depicted in Fig.~\ref{Fig:Intertwined_clusters_used}.

\section{Nearly synchronous clusters in each intertwined cluster set}
\label{sec:cs_intertwined}

\subsection{Condition for the stable, nearly synchronous CS of intertwined cluster sets}
\label{sec:stability}

For symmetry-induced CS pattern $\{C_m\}_{1\leq m \leq M}$,
we are interested in the emergence of a nearly synchronous CS of the pattern when all the oscillators are nearly identical.
Specifically, we consider Eq.~(\ref{eq:govern}) with 
$\boldsymbol{\upmu}_i = \overline{\boldsymbol{\upmu}}^m + \delta \boldsymbol{\upmu}_i$ for $i \in C_m$
with $||\delta \boldsymbol{\upmu}_i|| \ll 1$,
where $\overline{\boldsymbol \upmu}^m = (1/|C_m|)\sum_{i \in C_m} {\boldsymbol \upmu}_i$
is the average value of ${\boldsymbol \upmu}_i$ over the oscillators belonging to $C_m$.
Here, $||\delta \boldsymbol{\upmu}_i||$ denotes the Euclidean norm of $\delta \boldsymbol{\upmu}_i$, and
$|C_m|$ denotes the number of oscillators belonging to $C_m$.

Standard deviation of the states of the oscillators belonging to $C_m$, which is denoted by $\sigma_m$, is given by
\begin{equation}
\sigma_m(t)
=\sqrt{\frac{1}{|C_m|}\sum_{i \in C_m}||\delta\bold{x}_i(t)||^2},
\label{eq:std_original}
\end{equation}
where $\delta{\bold x}_i={\bold x}_i-\overline{\bold x}^m$ is the deviation of
${\bold x}_i$ from the average trajectory of $C_m$, which is
$\overline{\bold x}^m(t)=(1/|C_m|)\sum_{i \in C_m}{\bold x}_i(t)$. 
$\sigma_m$ can be written in a different form using 
the orthonormal set of {\it cluster-based} vectors denoted by
$\{{\bold u}^{(m)}_{\kappa}\}_{1 \leq \kappa \leq |C_m|}$.
Specifically, $u^{(m)}_{\kappa i} = 0$ if $i \notin C_m$ for each $N$-dimensional unit vector ${\bold u}^{(m)}_{\kappa}$.
We first define ${\bold u}^{(m)}_1$ as the unit vector whose nonzero elements are $1/\sqrt{|C_m|}$ constantly.
Therefore, ${\bold u}^{(m)}_1$ is the unit vector parallel to the synchronization manifold for $C_m$.
The other mutually orthogonal unit vectors $\{{\bold u}^{(m)}_{\kappa}\}_{2 \leq \kappa \leq |C_m|}$ 
are also orthogonal to ${\bold u}^{(m)}_1$, such that $\{{\bold u}^{(m)}_{\kappa}\}_{2 \leq \kappa \leq |C_m|}$ 
span the $(|C_m|-1)$-dimensional subspace transverse to the synchronization manifold for $C_m$.
For the set of unit vectors, we define the cluster-based coordinate system by
$\{\boldsymbol{\upeta}^{(m)}_{\kappa}~|~1 \leq \kappa \leq |C_m|,~1 \leq m \leq M\}$,
where $\boldsymbol{\upeta}^{(m)}_{\kappa} = \sum_{i \in C_m} u^{(m)}_{\kappa i}\delta {\bold x}_i$.

Using the new coordinate system $\{\boldsymbol{\upeta}^{(m)}_{\kappa}\}$, $\sigma_m$ is rewritten as
\begin{equation}
\sigma_m(t)
=\sqrt{\frac{1}{|C_m|}\sum_{\kappa=2}^{|C_m|}||\boldsymbol{\upeta}^{(m)}_{\kappa}(t)||^2},
\label{eq:std}
\end{equation}
where we use $\delta{\bold x}_i=\sum_{\kappa=1}^{|C_m|}u^{(m)}_{\kappa i}\boldsymbol{\upeta}^{(m)}_{\kappa}$
for $i \in C_m$
with $\sum_{i \in C_m}u^{(m)}_{\kappa i}u^{(m)}_{\kappa' i} = \delta_{\kappa\kappa'}$ by the orthogonality
between ${\bold u}^{(m)}_{\kappa}$ and ${\bold u}^{(m)}_{\kappa'}$ for $\kappa \neq \kappa'$, and
$\boldsymbol{\upeta}_1^{(m)}=(1/\sqrt{|C_m|})\sum_{i \in C_m}\delta {\bold x}_i = 0$ by the previous definition 
$\delta {\bold x}_i = {\bold x}_i - \overline{\bold x}^m$.
We remark that ${\{{\boldsymbol \upeta}^{(m)}_{\kappa}\}}_{2 \leq \kappa \leq |C_m|}$ 
determine the standard deviation of the oscillator states of $C_m$.

We analyze the dynamics of $\sigma_m$ using 
the variational equation of Eq.~(\ref{eq:govern}) along $\bold{x}_i = \overline{\bold x}^m$ for $i \in C_m$. 
This variational equation can be obtained by inserting 
$\bold{x}_i = \overline{\bold{x}}^m+\delta \bold{x}_i$ with 
$\boldsymbol{\upmu}_i = \overline{\boldsymbol{\upmu}}^m + \delta \boldsymbol{\upmu}_i$ into Eq.~(\ref{eq:govern}) 
for $i \in C_m$ as
\begin{eqnarray}
\delta \dot{\bold x}_i(t) &=& D_{\bold x}{\bold F}(\overline{\bold x}^m(t), \overline{\boldsymbol \upmu}^m)
\delta {\bold x}_i(t)
+ D_{\boldsymbol \upmu}{\bold F}(\overline{\bold x}^m(t), \overline{\boldsymbol \upmu}^m)\delta \boldsymbol{\upmu}_i
\notag \\
&+&\sigma \sum_{m'=1}^M\sum_{j \in C_{m'}} D_{\bold x}{\bf H}(\overline{\bf x}^{m'}(t))A_{ij}\delta {\bf x}_j(t)
\notag \\
&-&\frac{\sigma}{|C_m|}\sum_{i \in C_m}\sum_{m'=1}^M \sum_{j \in C_{m'}} D_{\bf x}{\bf H}
(\overline{\bf x}^{m'}(t))A_{ij}\delta{\bf x}_j(t), \notag \\
\label{eq:variation}
\end{eqnarray}
where $D_{\bold x}$ and $D_{\boldsymbol \upmu}$ denote the partial derivatives of each function
with respect to ${\bf x}$ and ${\boldsymbol {\upmu}}$, respectively.
Here, we use a Taylor expansion in the right hand side of Eq.~(\ref{eq:govern}) 
at $(\overline{\bold x}^m, \overline{\boldsymbol{\upmu}}^m)$
up to the linear order of $\delta \bold{x}_i$ and $\delta \boldsymbol{\upmu}_i$,
and obtain Eq.~(\ref{eq:variation}) by 
$\delta\dot{\bold x}_i = \dot{\bold x}_i - (1/|C_m|)\sum_{i \in C_m}\dot{\bold x}_i$~\cite{sorrentino2016}.

From now on, we consider a set of intertwined clusters $\{C_m\}_{1 \leq m \leq M'} \subseteq \{C_m\}_{1 \leq m \leq M}$
(after renumbering clusters as needed), where $M'$ is the number of clusters in the set of intertwined clusters.
Using the new coordinate system $\{{\boldsymbol \upeta}^{(m)}_{\kappa}\}$, Eq.~(\ref{eq:variation}) is rewritten as
\begin{equation}
\dot{\boldsymbol{\upeta}}^{(m)}_{\kappa}(t)=\sum_{m'=1}^{M'}\sum_{\kappa'=2}^{|C_{m'}|}
\bold{J}^{(mm')}_{\kappa \kappa'}(t){\boldsymbol{\upeta}}^{(m')}_{\kappa'}(t)
+{\bf b}^{(m)}_{\kappa}(t)
\label{eq:variation_transverse}
\end{equation}
for $2 \leq \kappa \leq |C_m|$, which determine the standard deviation of the oscillator states of $C_m$ by Eq.~(\ref{eq:std}), where
\begin{eqnarray}
\bold{J}^{(mm')}_{\kappa \kappa'}(t)&=&D_{\bold x}{\bold F}(\overline{\bold x}^m(t), \overline{\boldsymbol \upmu}^m)
\delta_{mm'}\delta_{\kappa\kappa'} \notag
\\&+&\sigma D_{\bold x}{\bold H}(\overline{\bold x}^{m'}(t))B^{(mm')}_{\kappa \kappa'} \notag
\end{eqnarray}
with
\begin{equation}
B^{(mm')}_{\kappa \kappa'}=\sum_{i \in C_m}\sum_{j \in C_{m'}}
u^{(m)}_{\kappa i}A_{ij}u^{(m')}_{\kappa' j} \notag
\end{equation}
and
\begin{equation}
\bold{b}^{(m)}_{\kappa}(t)=D_{\boldsymbol{\upmu}}{\bold F}
(\overline{\bold x}^m(t), \overline{\boldsymbol {\upmu}}^m)\sum_{i \in C_m}u^{(m)}_{\kappa i}\delta{\boldsymbol {\upmu}}_i \notag
\end{equation}
(see Appendix for derivation).

It has been shown that no choice of cluster-based coordinates $\{{\boldsymbol \upeta}^{(m)}_{\kappa}\}$
can make $B^{(mm')}_{\kappa \kappa'}=0$
for all of the pairs $\{(\kappa, \kappa')~|~2 \leq \kappa \leq |C_m|,~2 \leq \kappa' \leq |C_{m'}|\}$ 
in an arbitrary $C_{m'} \in \{C_{m'}\}_{1\leq {m' \neq m} \leq M'}$~\cite{yscho_prl_2017}.
Therefore, the standard deviation of the oscillator states of $C_m$ is coupled with that of all other clusters
in the same intertwined set. This means that the nearly synchronous clusters of each intertwined cluster set
should be formed or broken at the same time, such that we may
demonstrate the condition for the stable, nearly synchronous CS of each intertwined cluster set altogether.

We use $K$ to denote the number of dimensions of the subspace
transverse to the CS manifold for the set of intertwined clusters $\{C_m\}_{1 \leq m \leq M'}$, such that $K = \sum_{m=1}^{M'} (|C_m|-1)$.
We define $K \times N$ matrix $\bold U$ by
\begin{equation}
\bold{U}=[{\bf u}^{(1)}_{2},...,{\bf u}^{(1)}_{|C_1|},...,{\bf u}^{(M')}_{2},...,{\bf u}^{(M')}_{|C_{M'}|}]^{\top},
\end{equation}
where $\bold{U}^{\top}=\bold{U}^{-1}$ by the mutual orthogonality of the unit vectors
$\{{\bf u}^{(m)}_{\kappa}~|~2 \leq \kappa \leq |C_m|,~1 \leq m \leq M'\}$.

Then, Eq.~(\ref{eq:variation_transverse}) is rewritten in matrix form using $Kn \times 1$ matrix 
$\boldsymbol{\upeta}=
[{\boldsymbol \upeta}^{(1)\top}_{2},...,{\boldsymbol \upeta}^{(1)\top}_{|C_1|}
,...,{\boldsymbol \upeta}^{(M')\top}_{2},...,{\boldsymbol \upeta}^{(M')\top}_{|C_{M'}|}]^{\top}$ as
\begin{equation}
\dot{\boldsymbol{\upeta}}(t)={\bold J}(t){\boldsymbol{\upeta}}(t) + {\bold b}(t),
\label{eq:variational_matrix}
\end{equation}
where $Kn \times Kn$ matrix $\bold{J}$ is defined by
\begin{eqnarray}
\bold{J}(t)&=&\sum_{m'=1}^{M'}({\bold U}{\bold E}^{(m')}{\bold U}^{\top})
\otimes D_{\bold x}{\bold F}(\overline{\bf x}^{m'}(t), \overline{\boldsymbol{\upmu}}^{m'}) \notag \\ 
&+&\sigma(\bold{B}\otimes{\bold I}_n)\Bigg[\sum_{m'=1}^{M'}
({\bold U}{\bold E}^{(m')}{\bold U}^{\top})\otimes D_{\bold x}{\bold H}
(\overline{\bf x}^{m'}(t))\Bigg] \notag
\end{eqnarray}
for $K \times K$ matrix $\bold{B}={\bold U}{\bold A}{\bold U}^{\top}$. The $N \times N$ 
diagonal matrix ${\bf E}^{(m')}$, whose components are
\begin{flalign}
E^{(m')}_{ii}=
\begin{cases}
1 & \text{if}~~i \in C_{m'}, \\
0 & {\textrm{otherwise}},
\end{cases} \notag
\end{flalign}
and $Kn \times 1$ matrix ${\bf b}(t)$ is defined by
\begin{equation}
\bold{b}(t)=\Bigg[\sum_{m'=1}^{M'}({\bf U}{\bf E}^{(m')})\otimes 
D_{\boldsymbol {\upmu}}{\bf F}(\overline{\bold x}^{m'}(t), \overline{\boldsymbol{\upmu}}^{m'})\Bigg]
\delta{\boldsymbol{\upmu}} \notag
\end{equation}
for $Nq \times 1$ matrix 
$\delta \boldsymbol{\upmu} = [\delta{\boldsymbol \upmu}^{\top}_1,...,\delta{\boldsymbol \upmu}^{\top}_N]^{\top}$.
We note that $Kn$ is the number of dimensions of state space 
transverse to the CS manifold for the intertwined cluster set and $Nq$
is the number of dimensions of internal parameter space of whole oscillators.

We now solve the nonhomogeneous linear system in Eq.~(\ref{eq:variational_matrix}) 
for the time dependent $Kn \times Kn$ matrix ${\bold J}(t)$ and $Kn \times 1$ 
matrix ${\bold b}(t)$. 
We first assume that the largest Lyapunov exponent 
associated with the homogeneous part of Eq.~(\ref{eq:variational_matrix}),
$\dot{\boldsymbol{\upeta}}(t)={\bold J(t)}{\boldsymbol{\upeta}}(t)$, is negative. 
Under this assumption, the solution of the homogeneous part of
Eq.~(\ref{eq:variational_matrix}) is given by 
$\boldsymbol{\upeta}^*(t)=\boldsymbol{\Phi}(t, \tau)\boldsymbol{\upeta}^*(\tau)$
(i.e. $\dot{\boldsymbol{\upeta}}^*(t)=\bold{J}(t)\boldsymbol{\upeta}^*(t)$)
for $Kn \times Kn$ fundamental transition matrix $\boldsymbol{\Phi}(t, \tau)$, thereby satisfying
$||\boldsymbol{\Phi}(t, \tau)|| \leq \gamma e^{-\lambda(t-\tau)}$ for positive constants $\gamma$ and $\lambda$~\cite{takashi_nearly_identical}.
Then, the solution for $\boldsymbol{\upeta}(t)$ of Eq.~(\ref{eq:variational_matrix}) is given by
\begin{equation}
\boldsymbol{\upeta}(t)=\int_{0}^{t}\boldsymbol{\Phi}(t, \tau)\bold{b}(\tau)d\tau
\label{eq:variational_solution}
\end{equation}
as $t \rightarrow \infty$~\cite{perko_1996, rugh_1996}.

As a result, the nearly synchronous CS of each intertwined cluster set is stable when
$||\boldsymbol{\upeta}(t)||$ is bounded. This is guaranteed when
(i) the largest Lyapunov exponent associated with the homogeneous part of Eq.~(\ref{eq:variational_matrix})
is negative, and (ii) $||{\bold b}(t)||$ is bounded~\cite{takashi_nearly_identical}.

\subsection{Special case: Single cluster with parameter mismatch in a set of intertwined clusters}
\label{sec:special}

At first, Eq.~(\ref{eq:variational_solution}) can be expressed component-wise as
\begin{equation}
{\boldsymbol{\upeta}}^{(m)}_{\kappa}(t)
=\sum_{m'=1}^{M'}\sum_{\kappa'=2}^{|C_{m'}|}
\bigg[\int_{0}^{t}\boldsymbol{\Phi}^{(mm')}_{\kappa\kappa'}(t, \tau)
\bold{b}^{(m')}_{\kappa'}(\tau)d\tau\bigg]
\label{eq:variational_solution_component}
\end{equation}
$(2 \leq \kappa \leq |C_m|,~1 \leq m \leq M')$, where 
\begin{eqnarray}
&&\boldsymbol{\Phi}^{(mm')}_{\kappa\kappa'}(t, \tau)= \notag \\
&&\Big[\big(\bold{u}^{(m)\top}_{\kappa}{\bold E}^{(m)}{\bold U}^{\top}\big)
\otimes{\bold I}_n \Big]\boldsymbol{\Phi}(t,\tau)
\Big[\big(\bold{u}^{(m')\top}_{\kappa'}{\bold E}^{(m')}{\bold U}^{\top}\big)
\otimes {\bold I}_n\Big]^{\top} \notag.
\end{eqnarray}
In other words, $\boldsymbol{\Phi}^{(mm')}_{\kappa\kappa'}$ is the $n \times n$ block of
$\boldsymbol{\Phi}$ at $((m, \kappa), (m', \kappa'))$, where
$(m, \kappa)$ denotes the location of the column for 
${\bold u}^{(m)}_{\kappa}$ in ${\bold U}^{\top}$.

We now consider a special intertwined cluster set in which only one cluster, 
$C_{\widetilde{m}} \in \{C_m\}_{1\leq m \leq M'}$, is composed of nearly identical oscillators 
while every other cluster in the set is composed of identical oscillators 
(i.e. $\delta{\boldsymbol \upmu}_i = 0$ if $i \notin C_{\widetilde{m}}$).
We then regard the nearly synchronous CS of this intertwined cluster set as stable.
Under this circumstance, Eq.~(\ref{eq:variational_solution_component}) has the form
\begin{equation}
\boldsymbol{\upeta}^{(m)}_{\kappa}(t)=
\sum_{\kappa'=2}^{|C_{\widetilde{m}}|}
\bigg[\int_{0}^{t}\boldsymbol{\Phi}^{(m\widetilde{m})}_{\kappa\kappa'}(t, \tau)
\bold{b}^{(\widetilde{m})}_{\kappa'}(\tau)d\tau\bigg]
\label{eq:variational_solution_component_single}
\end{equation}
$(2 \leq \kappa \leq |C_m|,~1 \leq m \leq M')$.

We first assume that average trajectory $\overline{\bold x}^m(t)$ 
is close to quotient network dynamics ${\bold s}_m(t)$ of Eq.~(\ref{eq:govern_CS}) when
$||\delta \boldsymbol{\upmu}_i|| \ll 1$ for $\forall i$~\cite{sorrentino2016, bubbling_attractor1, bubbling_attractor2}.
Under this assumption, $\boldsymbol{\Phi}^{(m\widetilde{m})}_{\kappa\kappa'}$ is insensitive to variations in 
$\delta {\boldsymbol \upmu}_i$.
Then, if the magnitude of parameter mismatch of $C_{\widetilde{m}}$ 
is scaled by factor $c$ as $\delta {\boldsymbol{\upmu}}_i \rightarrow c\delta{\boldsymbol{\upmu}}_i$
for $i \in C_{\widetilde{m}}$
(i.e. ${\bold b}^{(\widetilde{m})}_{\kappa'} \rightarrow c {\bold b}^{(\widetilde{m})}_{\kappa'}$,~$2 \leq \kappa' \leq |C_{\widetilde{m}}|$), then
$\boldsymbol{\upeta}^{(m)}_{\kappa} \rightarrow c\boldsymbol{\upeta}^{(m)}_{\kappa}$
$(2 \leq \kappa \leq |C_m|,~1 \leq m \leq M')$
according to Eq.~(\ref{eq:variational_solution_component_single})
such that the standard deviation of every cluster in the intertwined cluster set $\{\sigma_m\}_{1 \leq m \leq M'}$ is scaled by the common factor $c$
following Eq.~(\ref{eq:std}).
This result implies that parameter mismatch in a single cluster can break the
synchronization of every other cluster in the same intertwined cluster set.
One example for this special case is presented in Sec.~\ref{sec:example}.

\section{Example}
\label{sec:example}

We apply the theoretical framework demonstrated in Sec.~\ref{sec:cs_intertwined}
to $y$-coupled R\"ossler oscillators in the network depicted in Fig.~\ref{Fig:Intertwined_clusters_used}.
For the state of each oscillator $\bold{x}_i=(x_i,y_i,z_i)^{\top}$, we consider
$\bold{F}({\bold x}_i, a_i)=(-y_i-z_i,~x_i+a_iy_i,~0.2+z_i(x_i-7))^{\top}$ and ${\bold H}({\bold x}_i)=(0, y_i, 0)^{\top}$. 
Therefore, the governing equation of this system is given by
\begin{eqnarray}
\dot{x}_i(t)&=& -y_i(t)-z_i(t), \notag \\
\dot{y}_i(t)&=& x_i(t)+a_iy_i(t)+\sigma\sum_{j=1}^N A_{ij} y_j(t), \notag \\
\dot{z}_i(t)&=& 0.2 + z_i(t)(x_i(t)-7).
\label{eq:example_governeq}
\end{eqnarray}
Here, we examine the CS pattern $\{C_1, C_2, C_3\}$ in Fig.~\ref{Fig:Intertwined_clusters_used}, 
where we assume that the two oscillators in $C_1$ are nearly identical with $a_1=a+\delta a$ and $a_2=a-\delta a$ for $\delta a \ll 1$,
while the other three oscillators in the network are identical 
with $a_i=a$ $(i=3,4,5)$. We note that $\overline{a}^m=a$ $(m=1,2,3)$.

\begin{figure}[t!]
\includegraphics[width=0.9\linewidth]{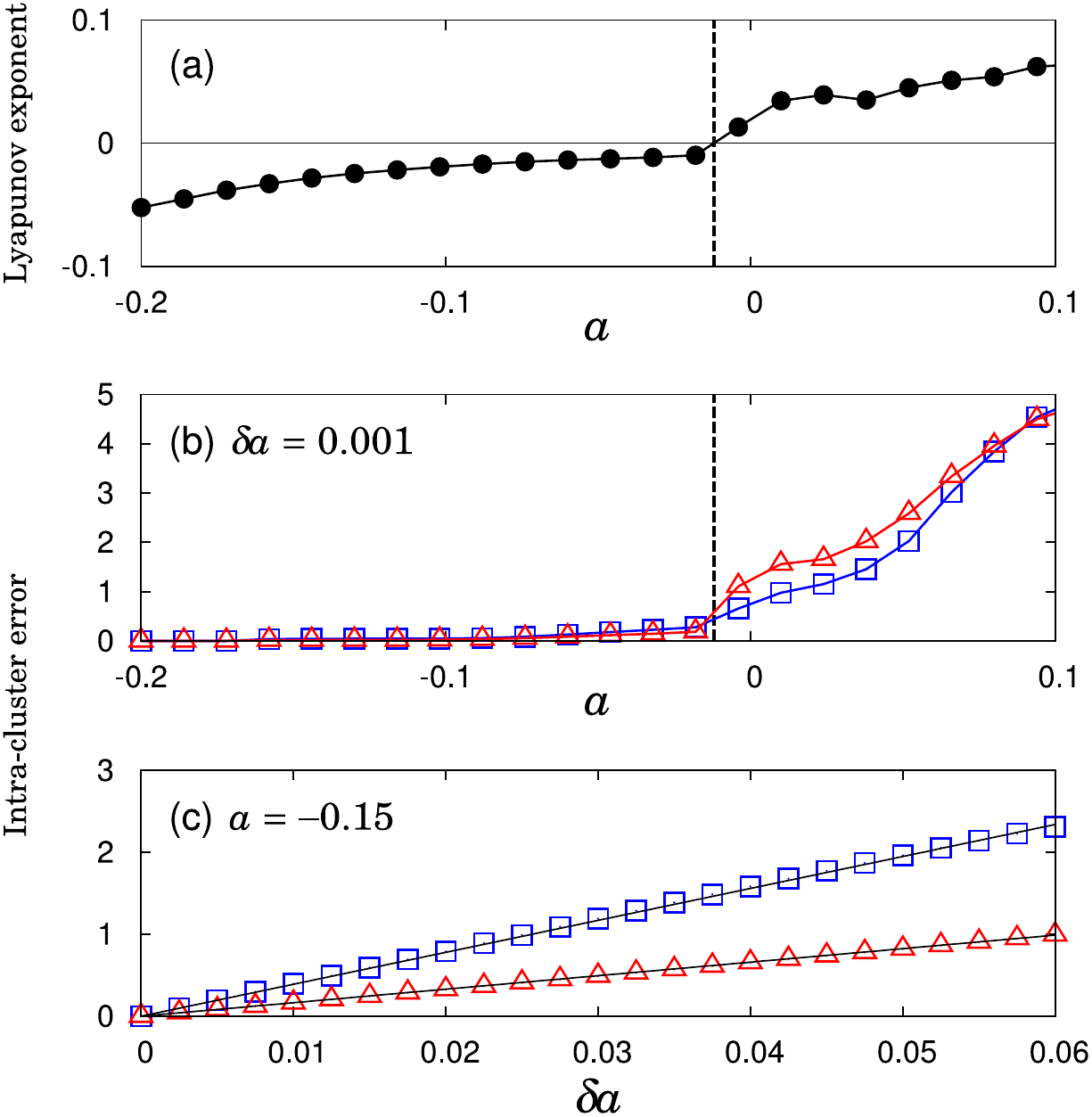}
\caption{Dynamics of $y$-coupled R\"ossler oscillators
for $\sigma=0.1$ in the network depicted in Fig.~\ref{Fig:Intertwined_clusters_used}.
(a) Numerically measured $\Lambda(a)$ $(\CIRCLE)$ as a function of $a$.
Nearly synchronous CS of the two clusters becomes unstable as $a$ exceeds the value of the vertical dashed line.
(b) For fixed $\delta a=10^{-3}$, both $\varOmega_1$ $(\square)$ and $\varOmega_2$ $(\triangle)$ drastically increase as
$a$ exceeds the value of the vertical dashed line. (c) For fixed $a=-0.15$, 
both $\varOmega_1$ $(\square)$ and $\varOmega_2$ $(\triangle)$ increase linearly with different slopes
as $\delta a$ increases. 
(a--c) To obtain each data point, we numerically integrate up to $T = 10^5$.}   
\label{Fig:3D_Rossler_y_twoISCset}
\end{figure}

As mentioned in Sec.~\ref{sec:intertwined} and Fig.~\ref{Fig:Intertwined_clusters_used}, 
$\{C_1,C_2\}$ is a set of intertwined clusters.
Stability of the nearly synchronous states of the two clusters is intertwined by
${\boldsymbol \upeta}=({\boldsymbol \upeta}^{(1)\top}_2, {\boldsymbol \upeta}^{(2)\top}_2)^{\top}$ 
in Eq.~(\ref{eq:variational_matrix}), where

\begin{equation}
{\bold J}(t)=
\begin{pmatrix}
0 & -1 & -1 & 0 & 0 & 0 \\
1 & a & 0 & 0 & \sigma & 0 \\
\overline{z}^1(t) & 0 & \overline{x}^1(t)-7 & 0 & 0 & 0 \\
0 & 0 & 0 & 0 & -1 & -1 \\
0 & \sigma & 0 & 1 & a & 0 \\
0 & 0 & 0 & \overline{z}^2(t) & 0 & \overline{x}^2(t) - 7
\end{pmatrix},
\end{equation}
and
\begin{equation}
{\bold b}(t)=-\sqrt{2}\delta a
\begin{pmatrix}
0 \\
\overline{y}^1(t) \\
0 \\
0 \\
0 \\
0
\end{pmatrix}.
\end{equation}

We numerically estimate the largest Lyapunov exponent associated with 
$\dot{\boldsymbol \upeta}(t) = \bold{J}(t){\boldsymbol \upeta}(t)$. 
To measure the exponent, we assume that average trajectory $\overline{\bold x}^m(t)$ for $\delta a \ll 1$ 
is close to quotient network dynamics ${\bold s}_m(t)$ of Eq.~(\ref{eq:govern_CS}) with $\overline{\boldsymbol{\upmu}}=a$
(i.e. $\delta a = 0$)~\cite{sorrentino2016, bubbling_attractor1, bubbling_attractor2}.
Then, we numerically integrate Eq.~(\ref{eq:govern_CS}) for this system, and
use $\overline{\bold x}^m(t) = {\bold s}_m(t)$ to integrate
$\dot{\boldsymbol \upeta}(t) = \bold{J}(t){\boldsymbol \upeta}(t)$ numerically. 
Finally, we measure the largest Lyapunov exponent by obtaining
$\Lambda(a)=(1/T)\textrm{ln}\big(||\boldsymbol{\upeta}(T)||/||\boldsymbol{\upeta}(0)||\big)$ for $T \gg 1$.  
To discard the initial transient, we numerically integrate Eq.~(\ref{eq:govern_CS}) 
for the time duration $T$ before obtaining the initial state $\bold{s}_m(0)$,
where each component of ${\bold s}_m(-T)$ is taken uniformly at random within the interval [-1, 1].
Estimated $\Lambda(a)$ for various values of $a$ is shown in Fig.~\ref{Fig:3D_Rossler_y_twoISCset}(a).

We want to show that both $C_1$ and $C_2$ can be nearly synchronous 
in the range of $a$ for $\Lambda(a) < 0$. 
For this purpose, we measure the {\it intra-cluster errors}, $\varOmega_1, \varOmega_2$, of the two clusters as given by
\begin{equation}
\varOmega_m=\frac{1}{T}\int_{0}^T\sigma_m(t)dt
\label{eq:intra_error}
\end{equation}
for $T \gg 1$.
Specifically, we numerically integrate Eq.~(\ref{eq:example_governeq}) directly to calculate $\sigma_m(t)$ using Eq.~(\ref{eq:std_original}).
To discard the initial transient, we numerically integrate Eq.~(\ref{eq:example_governeq}) 
for the time duration $T$ before obtaining the initial state $\bold{x}_i(0)$,
where each component of ${\bold x}_i(-T)$ is taken uniformly at random within the interval [-1, 1].

For a fixed $\delta a \ll 1$, we find that  
$\varOmega_1, \varOmega_2$ are small in the range of $a$ for $\Lambda(a)<0$,
while they are large in the range of $a$ for $\Lambda(a)>0$, as shown in Fig.~\ref{Fig:3D_Rossler_y_twoISCset}(b). 
We then verify that $||\bold{b}(t)||=\sqrt{2}\delta a||\overline{y}^1(t)||$ is also bounded for $0 \leq t \leq T$ when $\Lambda(a)<0$, which allows us to confirm that the nearly synchronous CS of the intertwined cluster set is stable when $\Lambda(a)<0$.

This system is an example of the special case discussed in Sec.~\ref{sec:special}.
Here, among the intertwined cluster set $\{C_1, C_2\}$,
$C_1$ is composed of two nearly identical oscillators with $a_1 = a+\delta a$ and $a_2 = a-\delta a$ for $\delta a \ll 1$,
while $C_2$ is composed of two identical oscillators with $a_3 = a_4 = a$.
As discussed in Sec.~\ref{sec:special}, both $\varOmega_1$ and $\varOmega_2$
increase linearly as $\delta a$---the magnitude of $C_1$ parameter mismatch---increases 
for a fixed $a$ when the nearly synchronous CS of the intertwined cluster set is stable (Fig.~\ref{Fig:3D_Rossler_y_twoISCset}(c)). 
This result demonstrates that synchronization between two identical oscillators within one cluster 
can be broken by two heterogeneous oscillators of the other cluster in the same intertwined cluster set.

\section{Discussion}
\label{sec:discussion}

Twisted states of identical oscillators, 
originally discovered in ring structures~\cite{sync_basin_2006, sync_basin_2017}, have been recently 
reported in square and cubic lattices~\cite{twisted_2018}.
In the twisted states of lattices, oscillators in each line are synchronized in square lattices
while oscillators in each plane are synchronized in cubic lattices. 
These states can be regarded as possible CS patterns of the given lattice, with 
each CS pattern of these states being a set of intertwined clusters
by the translational symmetry of the lattice. When the oscillators become heterogeneous,
every cluster in each CS pattern becomes nearly synchronous concurrently,  
which might be understood
using the theoretical framework established in this paper.

In the current work, we have considered nearly identical oscillators with time-independent internal parameters.
To describe more realistic systems, one might extend this work by considering systems with time-dependent nearly identical
internal parameters and entries of the coupling matrix~\cite{sorrentino_nearly_identical, sorrentino2016}. 
This extension would yield more fruitful results.

\section{acknowledgement}
This paper was supported by NRF Grant No. 2017R1C1B1004292.

\section*{Appendix: Derivation of  Eq.~(\ref{eq:variation_transverse})}
\label{appendix:variational_eq}

For each $C_m$ in the set of intertwined clusters $\{C_m\}_{1 \leq m \leq M'}$, we insert Eq.~(\ref{eq:variation})
into the right-hand side of $\dot{\boldsymbol{\upeta}}^{(m)}_{\kappa} = \sum_{i \in C_m} u^{(m)}_{\kappa i} \delta \dot{\bold x}_i$
for $2 \leq \kappa \leq |C_m|$,
such that
\begin{eqnarray}
&&\dot{\boldsymbol \upeta}^{(m)}_{\kappa}(t) = \sum_{i \in C_m}u^{(m)}_{\kappa i}\delta \dot{\bold x}_i(t) \notag \\
&=&D_{\bold x}{\bold F}(\overline{\bold x}^m(t), \overline{\boldsymbol \upmu}^m){\boldsymbol \upeta}^{(m)}_{\kappa}(t)
+D_{\boldsymbol{\upmu}}{\bold F}(\overline{\bold x}^m(t), \overline{\boldsymbol{\upmu}}^m)
\sum_{i \in C_m}u^{(m)}_{\kappa i}\delta {\boldsymbol \upmu}_i \notag \\
&+&\sigma \sum_{m'=1}^M \sum_{i \in C_m}\sum_{j \in C_{m'}} 
D_{\bold x}{\bold H}(\overline{\bold x}^{m'}(t))u^{(m)}_{\kappa i}A_{ij}\delta {\bold x}_j(t)
\notag \\ 
&-&\bigg(\sum_{i \in C_m} u^{(m)}_{\kappa i}\bigg)\frac{\sigma}{|C_m|}\sum_{k \in C_m}\sum_{m'=1}^{M}
\sum_{j \in C_{m'}}D_{\bold x}{\bold H}(\overline{\bold x}^{m'}(t))A_{kj}\delta {\bold x}_j(t). \notag \\
\label{eq:eq4_deriv1}
\end{eqnarray}
The last term of the right-hand side is deleted because $\sum_{i \in C_m} u^{(m)}_{\kappa i} = 0$ for $2 \leq \kappa \leq |C_m|$
(i.e. ${\bold u}^{(m)\top}_{\kappa}{\bold u}^{(m)}_{1}=\delta_{\kappa 1}$).
After inserting $\delta {\bold x}_j = \sum_{\kappa'=1}^{|C_{m'}|}u^{(m')}_{\kappa'j}\boldsymbol{\upeta}^{(m')}_{\kappa'}$ into the right-hand side of Eq.~(\ref{eq:eq4_deriv1}), it takes the form
\begin{eqnarray}
&&D_{\bold x}{\bold F}(\overline{\bold x}^m(t), \overline{\boldsymbol \upmu}^m){\boldsymbol \upeta}^{(m)}_{\kappa}(t)
+D_{\boldsymbol \upmu}{\bold F}(\overline{\bold x}^m(t), \overline{\boldsymbol \upmu}^m)
\sum_{i \in C_m}u^{(m)}_{\kappa i}\delta{\boldsymbol \upmu}_i  \notag \\
&+&\sigma \sum_{m'=1}^{M}\sum_{\kappa'=1}^{|C_{m'}|}
D_{\bold x}{\bold H}(\overline{\bold x}^{m'}(t))
\bigg(\sum_{i \in C_m}\sum_{j \in C_{m'}}
u^{(m)}_{\kappa i}
A_{ij}u^{(m')}_{\kappa' j}\bigg){\boldsymbol \upeta}^{(m')}_{\kappa'}(t) \notag \\
&=&D_{\bold x}{\bold F}(\overline{\bold x}^m(t), \overline{\boldsymbol \upmu}^m){\boldsymbol \upeta}^{(m)}_{\kappa}(t)
+D_{\boldsymbol \upmu}{\bold F}(\overline{\bold x}^m(t), \overline{\boldsymbol \upmu}^m)
\sum_{i \in C_m}u^{(m)}_{\kappa i}\delta{\boldsymbol \upmu}_i  \notag \\
&+&\sigma \sum_{m'=1}^{M'}\sum_{\kappa'=2}^{|C_{m'}|} D_{\bold x}{\bold H}(\overline{\bold x}^{m'}(t))
B^{(mm')}_{\kappa \kappa'}{\boldsymbol \upeta}^{(m')}_{\kappa'}(t) \notag \\
&=& \sum_{m'=1}^{M'}\sum_{\kappa'=2}^{|C_{m'}|} 
\Big[D_{\bold x}{\bold F}(\overline{\bold x}^m(t), \overline{\boldsymbol \upmu}^m)\delta_{mm'}\delta_{\kappa\kappa'} \notag \\
&+&\sigma D_{\bold x}{\bold H}(\overline{\bold x}^{m'}(t))B^{(mm')}_{\kappa\kappa'}\Big] 
{\boldsymbol \upeta}^{(m')}_{\kappa'}(t)
+D_{\boldsymbol \upmu}{\bold F}(\overline{\bold x}^m(t), \overline{\boldsymbol \upmu}^m)
\sum_{i \in C_m}u^{(m)}_{\kappa i}\delta{\boldsymbol \upmu}_i, \notag \\
\end{eqnarray}
where we used $B^{(mm')}_{\kappa\kappa'}=\sum_{i \in C_m}
\sum_{j \in C_{m'}}u^{(m)}_{\kappa i}A_{ij}u^{(m')}_{\kappa' j}$ with $B^{(mm')}_{\kappa 1}=0$ and
$B^{(mm')}_{\kappa \kappa'}=0$ for $M'+1 \leq m' \leq M$
(for $m'$ outside of the intertwined cluster set $\{C_m\}_{1 \leq m \leq M'}$)~\cite{yscho_prl_2017}.

\end{document}